\documentclass[pra, 10pt, twocolumn, tightenlines, longbibliography, aps,superscriptaddress,nofootinbib]{revtex4-2}

\usepackage[dvips]{graphics}
\usepackage{color}
\usepackage{float}
\usepackage{natbib}
\usepackage[dvipsnames]{xcolor}
\setcitestyle{numbers}
\setcitestyle{square}
\usepackage{graphicx}
\usepackage{bm}
\usepackage{amssymb}
\usepackage{amsmath}
\usepackage{mathtools}
\usepackage{dcolumn}  
\usepackage[colorlinks, linkcolor=Blue,citecolor=Blue,urlcolor=Blue]{hyperref}

\usepackage[all]{hypcap} 
\usepackage{physics}
\usepackage{enumerate}
\usepackage{braket}       
\usepackage{overpic}
\usepackage{amsfonts}
\usepackage{dsfont}
\usepackage[mathscr]{eucal}
\usepackage{hyperref}
\usepackage{setspace}
\usepackage{lipsum} 

\usepackage[normalem]{ulem} 

\begin{document}
\title{Proposals for realizing a Josephson diode in Atomtronic circuits}
\author{Nalinikanta Pradhan}
\affiliation{Department of Physics, Indian Institute of Technology, Guwahati 781039, Assam, India}

\author{Rina Kanamoto}
\affiliation{Department of Physics, Meiji University, Kawasaki, Kanagawa 214-8571, Japan}

\author{M. Bhattacharya}
\affiliation{School of Physics and Astronomy, Rochester Institute of Technology, 84 Lomb Memorial Drive, Rochester, New York 14623, USA}
\author{Pankaj Kumar Mishra}
\affiliation{Department of Physics, Indian Institute of Technology, Guwahati 781039, Assam, India}

\date{\today}

\begin{abstract} 

The Josephson diode, a non-reciprocal quantum element analogous to the familiar semiconductor \textit{p--n} junction diode, has been realized in solid-state systems but remains unexplored in tunable atomtronic circuits. In this work, we propose and numerically demonstrate the realization of the Josephson diode effect in an atomtronic circuit consisting of a ring-shaped Bose–Einstein condensate and with optical barriers serving as Josephson junctions. Our implementation of this macroscopic non-reciprocal quantum phenomenon is based on realizing the required inversion symmetry breaking through asymmetric barrier placement and an asymmetric alternating current (AC) drive, enabling position- and drive-tunable diode effects with efficiencies up to 15\% and 91\%, respectively. While standard time-of-flight absorption imaging can readily observe these effects, we employ cavity optomechanics for \textit{in situ}, real-time, and non-destructive measurements of the relevant condensate dynamics. Our results establish a highly tunable platform for nonreciprocal Josephson transport, opening avenues for diode-based neutral-atom technologies in future quantum circuits.

\end{abstract}


\maketitle

\textit{Introduction:} The semiconductor \textit{p--n} junction diode, a foundational element of modern electronics, serves as a classic example of a non-reciprocal circuit element, and finds application in signal processing, logic circuits, and sensors~\cite{sze2008semiconductor, sensorsPNdiode, mehta2022principles}. Inspired by the success of 
the \textit{p--n} junction diode, a key goal has been to realize a superconducting analogue that exhibits similar non-reciprocal behavior, but without the inherent dissipation due to resistance. This has led to  theoretical proposals for realizing the superconducting diode effect (SDE) in Josephson junctions, termed as the Josephson diode effect (JDE)~\cite{ProposedDiodePRL,Misaki2021theory, Zhang2022TheoryJDE, shaffer2025theoriessuperconductingdiodeeffects}. 


Experimentally, the SDE was initially observed in junction-free superconductors~\cite{ando2020observation,Itahashi2020,Miyasaka_2021,Kawarazaki_2022,narita2022field,lin2022zero,Du2024_sde,samanta2025fieldfreesuperconductingdiodeeffect,Chen2025IntrinsicSDE}, followed by its realization in Josephson junctions (JJs)~\cite{Turini2022JDE,diez2023symmetry,trahms2023diode,matsuo2023josephson}, and in superconducting quantum interference devices (SQUID)~\cite{Souto2022PRL, Paolucci2023, Ciaccia2023, li2024interfering, Coraiola2024Flux-tunable}. This effect requires the breaking of inversion symmetry (IS) and time-reversal symmetry (TRS) of the system, achieved through an external magnetic field or intrinsic spin-orbit coupling~\cite{Zhang2022TheoryJDE}. Remarkably, recent experiments have demonstrated field-free superconducting diodes that rely solely on IS breaking, achieving robust half-wave rectification over multiple cycles~\cite{wu2022field, Nagata2025PRL}. 
Theoretically, these effects have been widely explored using the Resistively and Capacitively Shunted Junction (RCSJ) model~\cite{Fominov2022AsymSQUID,wang2025rcsj}, Bogoliubov-de Gennes analysis~\cite{Li2024TB,Boruah2025,samanta2025,Sharma2025} and Green's function techniques~\cite{Shaffer2025Greenfunction}. A prominent physical platform for investigating these effects is the asymmetric SQUID, where a non-sinusoidal current-phase relationship (CPR) and external flux bias give rise to the non-reciprocal transport~\cite{Fulton1972, Cuozzo2024, Leblanc2025,Qi2025Non-Hermitian}. 

Beyond solid-state superconducting diodes, \textit{atomtronic circuits}, composed of Bose-Einstein condensates (BECs), offer a powerful and complementary platform for exploring the Josephson diode effect. They provide exceptional control over system geometry, atomic interactions, disorder, and synthetic gauge fields, allowing direct manipulation of coherent matter waves~\cite{Lewenstein2012UltracoldAtoms, Amico_2017Focus, Amico2021roadmap, Polo_2024}.
In typical atomtronic implementations, JJs are realized using tightly focused, repulsive optical barriers that separate a BEC into two or more weakly coupled regions~\cite{Henderson_2009Painting, wright2013driving, RyuPRL2013}. A bias current is generated by imposing a relative motion between the barrier and the condensate~\cite{levy2007acdc, RyuPRL2013, KSgan2025josephson2D}. The critical current that the barrier can support is determined by the current-chemical potential difference (I-V) relation, which is measured by counting the number of atoms in the two wells using absorption imaging~\cite{levy2007acdc,experiment1DJJ, bernhart2024observation, singh2024,KSgan2025josephson2D}. However, these traditional imaging techniques are destructive and require repeated preparation of the BEC. Recently, we have addressed this limitation by employing cavity optomechanics to infer the chemical potential difference in an atomtronic SQUID in a non-destructive, \textit{in situ}, real-time, single-shot manner~\cite{pradhan2025fractional}, offering a major advance over the absorption imaging-based methods.

In this Letter, we theoretically propose an atomtronic Josephson diode by breaking inversion symmetry in a ring BEC using two experimentally accessible configurations: (i) asymmetrically placed JJs, and (ii) a biharmonic ac drive. In the first configuration, asymmetrically placed junctions divide the ring into arms of unequal length. Under a bias current induced by the junction motion, the dynamics become direction-dependent, and the DC–AC transition occurs earlier when the junctions move toward the shorter arm. This results in nonreciprocal critical currents ($|I_{c+}| \neq |I_{c-}|$), thereby realizing a position-tunable Josephson diode. In the second configuration, biharmonic modulation of the junction positions, with a non-zero phase difference between the two tones, breaks the spatio-temporal symmetry of the system,  enabling a drive-tunable diode~\cite{Borgongino2025biharmonic,tsarev2025fractionalshapirostepsrsj}.

\begin{figure}[b]
\begin{center}
    \includegraphics[width=1.0\linewidth]{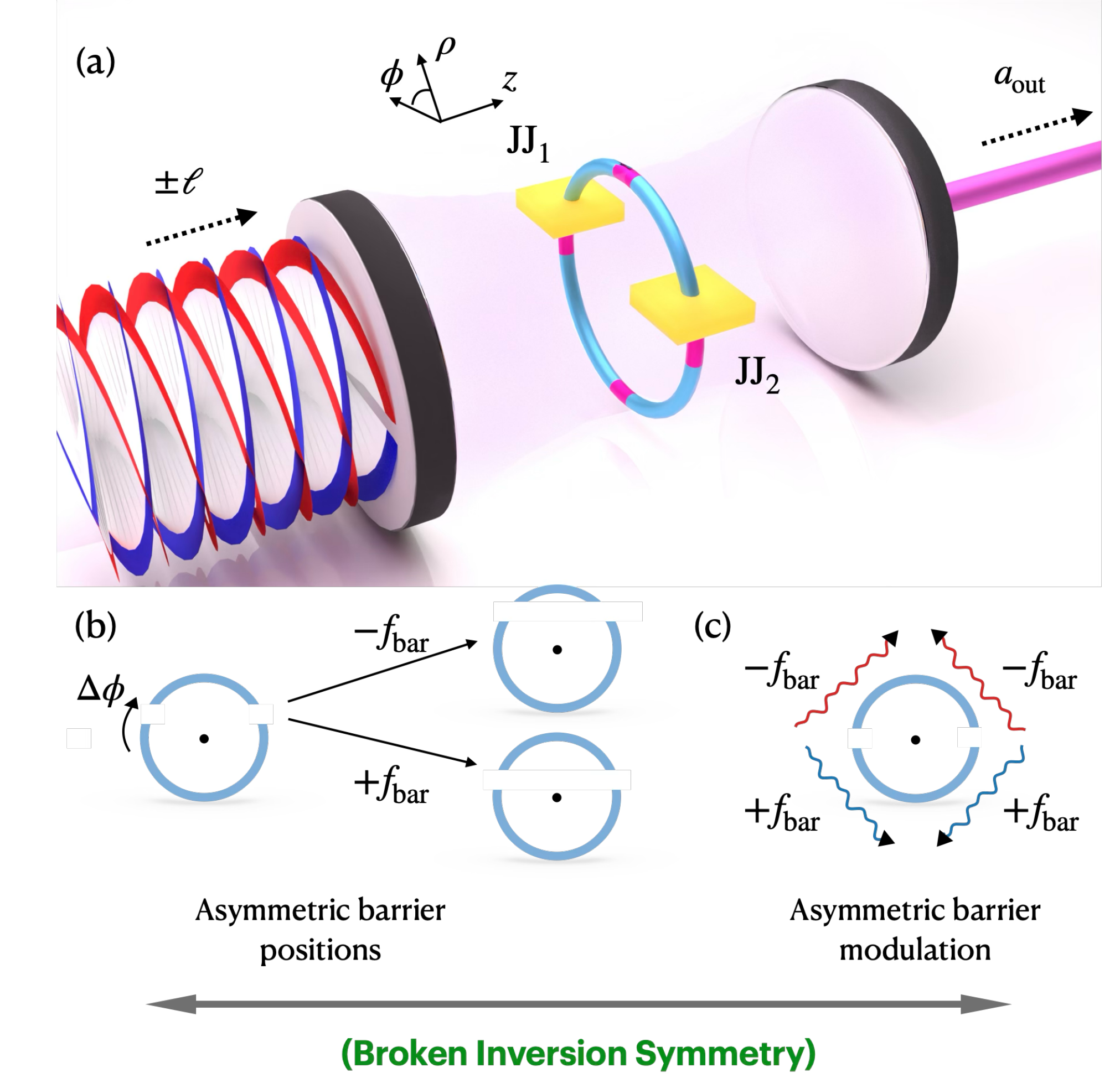} \\

\end{center}
\vspace{-12pt}
\caption{(a) Schematic diagram of the proposed experimental setup. A ring BEC is kept at the centre of an optical cavity that is driven by the superposition of two LG beams. The output optical field from the cavity is denoted by $a_{\mathrm{out}}$. (b) Schematic protocol to break the inversion symmetry of the condensate by placing the junction such that it divides the ring into two unequal arms. Here $\Delta\phi$ quantifies deviation from the diametrically opposite positions. (c) Schematic protocol to break the spatio-temporal symmetry by an asymmetric ac drive to the initially symmetric junctions. }

\label{fig:setup}
\end{figure}

\textit{Physical setup:} Our proposed experimental setup consists of a ring BEC of $^{23}$Na atoms, placed at the center of a high-finesse optical cavity, as shown in Fig.~\ref{fig:setup} (a). By denoting the radius of the ring trap as $R$, the radial and axial harmonic trapping frequencies as $\omega_\rho$ and $\omega_z$, respectively, the trapping potential can be expressed as $U(\rho, z) = \frac{1}{2}\, m \omega_\rho (\rho-R)^2 + \frac{1}{2}\, m \omega_z z^2$, where $m$ is the mass of each atom. As a result of tight confinement along the radial and axial directions, atoms remain frozen in the ground state along those directions throughout the evolution, and this makes the azimuthal degree of freedom along $\phi$ as the only relevant dynamical variable. Therefore, the system can be reduced to an effective one-dimensional description, which is accessible with current experimental capabilities~\cite{AmicoReview2024} and has been successfully applied to model experiments \cite{wright2013driving}, with the condition on the total number of atoms $N$ as~\cite{MorizotPRA2006,KumarPRL2021}
\begin{equation}
    N\lesssim \frac{4 \sqrt{\pi}R}{3a_{\mathrm{s}}} \bigg(\frac{\omega_\rho}{\omega_z}\bigg)^{\frac{1}{2}},
    \label{eq:1d_limit}
\end{equation}
where $a_\mathrm{s}$ is the $s$-wave scattering length of the atoms.


We analyze the Josephson dynamics by moving the JJs towards each other with velocity $f_{\mathrm{bar}}$ up to time $t_{\mathrm{bar}}$. This results in a DC supercurrent (AC current) if $f_{\mathrm{bar}}<f_{\mathrm{c}}$ ($f_{\mathrm{bar}}>f_{\mathrm{c}}$), where, $f_{\mathrm{c}}$ is the critical barrier velocity~\cite{RyuPRL2013,levy2007acdc}. To probe the Josephson oscillation frequency, we drive the cavity with a superposition of two Laguerre-Gaussian (LG) beams with orbital angular momentum (OAM) $\pm \ell \hbar$, which form an angular optical lattice potential about the cavity axis~\cite{NaidooAPB2012}. These beams are blue-detuned far from any atomic transition, and interact weakly with the condensate atoms. After the light-matter interaction, the cavity output field carries the signature of BEC dynamics (i.e., oscillation frequencies) and can be measured by the homodyne detection technique of cavity optomechanics~\cite{AspelmeyerRMP2014}.



\textit{Theoretical Model:} To numerically simulate the above physical system, we describe the azimuthal motion of BEC by a 1D Hamiltonian~\cite{KumarPRL2021,pradhan2024cavity, pradhan2024ring}
\begin{equation}
\begin{aligned}
    H&=\int_{0}^{2\pi}d\phi \;\Psi^{\dagger}(\phi)\Big[\mathcal{H}^{(1)}+\frac{g}{2}\Psi^{\dagger}(\phi)\Psi(\phi)\Big]\Psi(\phi)\\
    &-\hbar\Delta_{o}a^{\dagger}a+i\hbar\eta_{\mathrm{p}}(a^{\dagger}-a),\label{eq:Hamiltonian}
\end{aligned} 
\end{equation}
where, $\Psi(\phi)$ and $a$ are the bosonic and photonic operators that obey $[\Psi(\phi), \Psi^\dagger(\phi')] = \delta(\phi - \phi')$ and $[a, a^\dagger] = 1$, respectively. $\mathcal{H}^{(1)}$ represents the single-atom Hamiltonian: $\mathcal{H}^{(1)}=\frac{\hbar^2}{2mR^2}\left( -i \frac{d}{d\phi} - \Omega'\right)^2+\hbar U_{0}\cos^{2}(\ell\phi)a^{\dagger}a + V_{\mathrm{bar}}$. 
Here, the first term denotes the rotational kinetic energy of the atoms, the second the optical lattice potential with strength $U_0$ (See the supplementary material \cite{supplement} for a detailed derivation), and the third the optical barrier potentials, modelled as $V_{\mathrm{bar}}(\phi,\tau) =
\sum_{j=1}^{2} 
V_j \exp\!\left[-2\left(\tfrac{\phi-\phi_{b_j}(\tau)}{\sigma_j}\right)^{2}\right],$
with barrier height $V_j$, width $\sigma_j$ and barrier position $\phi_{b_j}$. The second term in Eq.~(\ref{eq:Hamiltonian}) refers to the two-body atomic interaction energy with strength $g$, and the last two terms denote the energy contributions from the cavity field and laser drive, respectively. $\Delta_{o}=\omega_{\mathrm{L}}-\omega_{0}$ denotes the drive-cavity detuning and $\eta_{\mathrm{p}}=\sqrt{P_{\mathrm{in}}\gamma_{0}/\hbar\omega_{0}}$ is the pump rate in terms of input optical power $P_{\mathrm{in}}$ and cavity decay rate $\gamma_{0}$.

Using a mean-field description and taking into account thermal fluctuations and laser shot noise, we describe the temporal evolution of the condensate wave function $\psi$ and the light field amplitude $\alpha$ as \cite{das2012winding,AspelmeyerRMP2014, KumarPRL2021, pradhan2024cavity, pradhan2025AndreevBashkin}
\begin{equation}
\begin{aligned}
    (i-\Gamma)\frac{d\psi}{d\tau} &= \biggl[\left( -i \frac{d}{d\phi} - \Omega'\right)^2 + \frac{U_0}{\omega_\beta}  |\alpha(\tau)|^2 \cos^2\left({\ell\phi}\right) \\ & + V_{\mathrm{bar}} (\phi,\tau)   + \mathcal{G} | \psi (\phi,\tau)|^2 \biggr] \psi + \xi(\phi,\tau),
    \label{Eq:bec}    
\end{aligned}
\end{equation}
\begin{equation}
\begin{aligned}
    i\frac{d\alpha}{d\tau} = \biggl\{ - \biggl[\Delta_c &- U_0 \langle \cos^{2}\left(\ell\phi\right)\rangle_{\tau} + i \frac{\gamma_{0}}{2}\biggr] \alpha  + i\eta_{\mathrm{p}} \biggr\} \omega_\beta^{-1} \\ &+ i \sqrt{\gamma_0} \omega_\beta^{-1} \alpha_{in}(\tau). 
    \label{Eq:cavity}
\end{aligned}
\end{equation}
using scaled energy $\hbar \omega_{\beta} = \hbar^2 / 2 m R ^2$, time $\tau = \omega_\beta t$, rotation rate $\Omega' = \Omega/2\omega_\beta$, atomic interaction strength $\mathcal{G} = g/ \hbar\omega_\beta$, and a phenomenological damping term $\Gamma$. Previously, a similar description has been successfully used to model BEC-cavity experiments, and they have shown excellent agreement with the mean-field simulation results~\cite{BrenneckeScience2008, kollar2017supermode}. Interaction with the external environment is taken into account through adding the stochastic terms to Eqs.~(\ref{Eq:bec}) and (\ref{Eq:cavity}) that denote the condensate's thermal fluctuations and the laser shot noise, respectively, and are described by delta-correlated white noise, with correlations~\cite{das2012winding,KumarPRL2021}
\begin{align} 
\langle\xi(\phi,\tau) \, \xi ^ *(\phi',\tau')\rangle  &= \frac{2\,\Gamma \,k_B \,T}{\hbar \omega_\beta} \, \delta(\phi - \phi') \, \delta (\tau - \tau'), \\ 
\langle \alpha_{in}(\tau) \, \alpha_{in}^ *(\tau') \rangle  &= \omega_\beta \, \delta (\tau - \tau'),
\end{align}
where $k_B$ is the Boltzmann constant and $T$ is the BEC temperature. 

\textit{Simulation details:} We generate the ground state of the condensate in the presence of optical barriers using the imaginary time evolution method~\cite{pradhan2024cavity}. For simulating the barrier movement step, their positions are modeled as $\phi_{b_{1,2}} = \pm(2 \pi f_{\mathrm{bar}} ) t$, in the rotating frame, up to time $t_{\mathrm{bar}}$. Subsequently we solve the coupled BEC-cavity equation (Eqs.~(\ref{Eq:bec}) and (\ref{Eq:cavity})) for a range of barrier velocities up to time $t_{\mathrm{sim}}$, which allows us to characterize the superfluid (dc) and dissipative (ac) regime from the power spectrum of phase quadrature of the cavity output field ($S(\omega)=\left|\mathrm{Im}\left[\alpha_{out}(\omega)\right]\right|^{2}$). The cavity output field $\alpha_{out} (t)$ is related to the intra-cavity field and optical fluctuations through $\alpha_{out} = -\alpha_{in} + \sqrt{\gamma_0}\alpha$ \cite{AspelmeyerRMP2014}. The cavity field is activated when the barrier motion ceases, ensuring that it does not capture any transient dynamics during barrier motion. This procedure yields the same results, though in real time, \textit{in situ}, and non-destructively, as the methods of detecting Josephson dynamics via absorption imaging of the condensate once the barrier motion stops~\cite{RyuPRL2013, ryu2020quantum, bernhart2024observation, KSgan2025josephson2D}. 



\begin{figure}[b]
\begin{center}
\includegraphics[width= 1\linewidth]{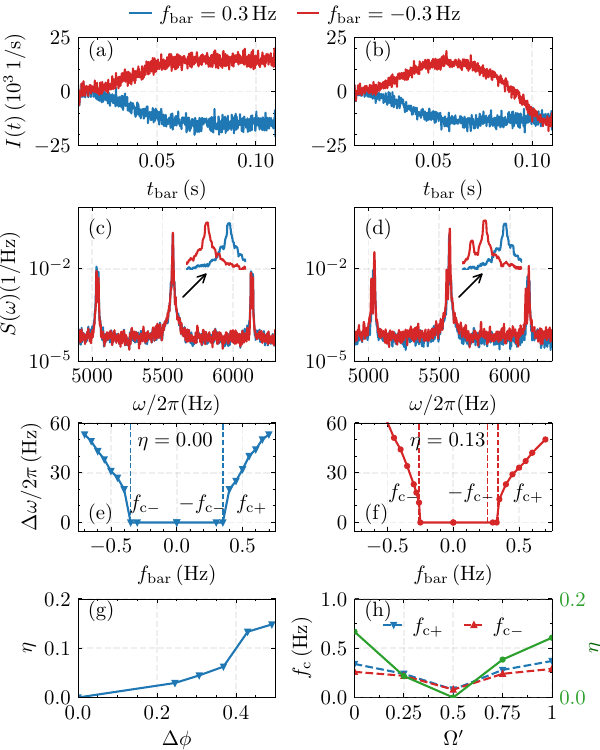} \\

\end{center}
\vspace{-12pt}
\caption{Position tunable diode: [(a)-(b)] Josephson tunneling current during the junction movement period with frequency $f_{\mathrm{bar}} = \pm 0.3$ Hz for (a) initially symmetric barriers ($\Delta \phi = 0$), (b) initially off-centered barriers ($\Delta \phi = 0.43$). [(c)-(d)] Power spectrum of phase quadrature of cavity output field corresponding to the case of (a) and (b), respectively. In the inset, the curves are deliberately displaced horizontally for clarity. [(e)-(f)] Magnitude of peak splitting in the cavity output spectrum versus barrier velocity for (e) $\Delta \phi = 0$ and (f) $\Delta \phi = 0.43$. The vertical dashed lines indicate the critical barrier velocity of the dc-to-ac Josephson transition. (g) Variation of diode efficiency $\eta$ with the amount of initial asymmetry $\Delta \phi$. (h) Variation of critical barrier velocity $f_{\mathrm{c}}$ and $\eta$ vs AQUID rotation rate $\Omega'$. The other parameters used are $N =  3700$, $R_0 = 4$ $\mu$m, $V_0 = 17.5\,\mu_0 = 25$ nK , $\sigma = 0.7 \, \zeta = 1.8\,\mu$m, $\zeta = 1/\sqrt{\mu_0/\hbar \omega_\beta}$, $t_{\mathrm{bar}}=0.1$ s, $T = 10$ nK, $P_{\mathrm{in}} = 10 \, \mathrm{fW}$, $\Gamma = 0.0001$, $ U_0 = 2\pi \times 212$ Hz, $\Tilde{\Delta} = \Delta_0 - U_0 N/2 = -2 \pi \times 173$ Hz, $\Delta_a = 2\pi \times 4.7$ GHz, $\omega_0 = 2\pi \times10^{15} $ Hz, $\omega_\rho = \omega_z = 2\pi \times 42$ Hz, $\gamma_{0} = 2\pi\times 2$ MHz, and $\Omega' = 0$.} 
\label{fig:fig2}
\end{figure}

\textit{Detection of the diode effect:} We next focus on achieving the position-tunable diode. Figs.~\ref{fig:fig2} (a)-(b) show the temporal evolution of the atomic tunneling current as the barriers move with velocities $\pm f_{\mathrm{bar}}$. When the barriers are symmetrically placed across the diameter of the ring ($\Delta \phi = 0$), equal and opposite barrier velocities generate equal and opposite DC supercurrents around the ring. However, for asymmetrical placed barriers (see Fig.~\ref{fig:setup} (b)), a positive bias (toward the longer arm), produces a dc current, while the same negative bias (toward the shorter arm), leads to oscillation in the current. This behaviour indicates that the system has entered into the dissipative regime for negative biasing, while it remains superfluid for positive biasing. In our setup, this discrepancy is detected via the power spectrum of the phase quadrature of the cavity output field $S(\omega)$, which characterizes the dc (ac) Josephson regime by zero (non-zero) splitting of all the side mode peaks, and the magnitude of splitting $\Delta \omega$ is the same as the Josephson oscillation frequency~\cite{pradhan2025fractional}. 


We have presented the cavity output spectrum $S(\omega)$ for the symmetric and asymmetric cases in Figs.\ref{fig:fig2} (c) and (d), respectively.
In the symmetric case, the spectrum remains the same and exhibits no splitting of side modes for both positive and negative current biases. However, in the asymmetric case, the spectrum shows side-mode splitting for negative biasing but not for positive biasing. This supports the findings from Figs.~\ref{fig:fig2} (b) that the dc current flows for only positive biasing for the choosen barrier velocity. We plot the magnitude of splitting $\Delta \omega$ versus barrier velocity $f_\mathrm{bar}$, a current (I)-voltage (V) curve analogue, for both cases in Figs.~\ref{fig:fig2} (e) and (f). The critical velocities in the positive and negative directions are denoted by $f_\mathrm{c+}$ and $f_\mathrm{c-}$, respectively, beyond which $\Delta \omega \neq 0$. For the symmetric case, the I-V curve is also symmetric, and the critical velocities in both directions are the same. In the other case, $f_\mathrm{c+}$ and $f_\mathrm{c-}$ differ by a non-zero amount, which signify that we achieve a diode-like behavior for particular barrier velocities ($|f_\mathrm{c-}| < |f_\mathrm{bar}| < |f_\mathrm{c+}|$), when the system remains in the superfluid regime (dc current) for positive biasing and in the resistive regime (ac current) for equal negative biasing. We quantify this difference in critical barrier velocities as the diode efficiency factor $\eta = \frac{|I_\mathrm{c+}| - |I_\mathrm{c-}|}{|I_\mathrm{c+}| + |I_\mathrm{c-}|} = \frac{|f_\mathrm{c+}| - |f_\mathrm{c-}|}{|f_\mathrm{c+}| + |f_\mathrm{c-}|}$~\cite{Souto2022PRL,Coraiola2024Flux-tunable,Ciaccia2023}.

\begin{figure}[!b]
\begin{center}
\includegraphics[width= 1\linewidth]{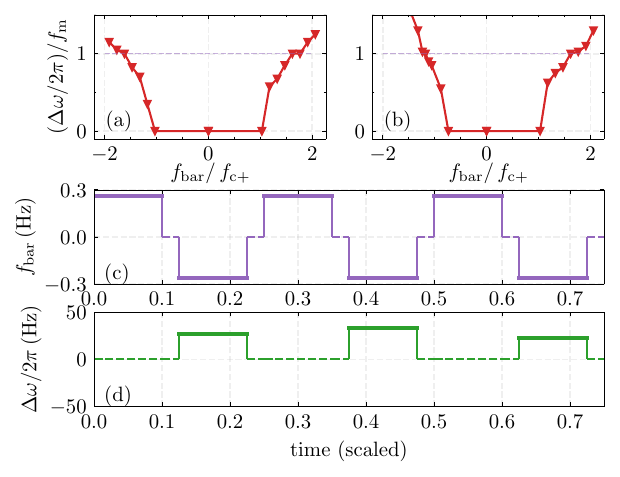} \\

\end{center}
\vspace{-12pt}
\caption{Applications of atomtronic Josephson diode: [(a)-(d)] Direction dependent modulation detector: Variation of the magnitude of peak splitting ($\Delta \omega$) versus barrier velocity ($f_{\mathrm{bar}}$) for $f_{\mathrm{m}} =40\,\mathrm{Hz},\,  \phi_\mathrm{m} = 0.06$, (a) $\Omega' = 0$, $\Delta\phi = 0$, (b) $\Omega' = 0$, $\Delta\phi = 0.43$. [(c)-(d)] Half-wave rectifier: (c) Three cycles of square wave of bias current that corresponds to $f_{\mathrm{bar}} = \pm 0.26$ Hz (lies in between $f_{\mathrm{c}+}$ and $f_{\mathrm{c}-}$). (d) Measured junction voltage ($\Delta \omega$). In the horizontal axes of (c) and (d), $t_{\mathrm{sim}}$ is scaled down to $t_{\mathrm{bar}}/4$ for a clear demonstration of current cycles. Here $t_{\mathrm{bar}} = 0.1$ s, $f_{\mathrm{m}} =0.0\,\mathrm{Hz},\,  \phi_\mathrm{m} = 0.0$, $\Omega' = 0$, and the other set of parameters are the same as used in Fig.~\ref{fig:fig2}.}
\label{fig:fig3}
\end{figure}

The physical origin of this diode effect is related to the way the phase difference across the barriers $\Phi = \Phi_1 - \Phi_2$ evolves when the barriers are in motion. Moving barriers develop a density imbalance leading to a non-zero chemical potential bias. For low barrier velocities, atoms tunnel through the barriers and cancel the bias. As a result, the phase difference $\Phi$ remains constant, giving rise to a dc supercurrent. At higher velocities, the density imbalance builds up so quickly ($\Delta \mu \neq 0$) that the phase difference cannot stay constant and starts running, $\frac{d\Phi}{dt} = -\frac{\Delta \mu}{\hbar}$ and the current starts oscillating as $I = I_c \sin(\Phi)$. When the barriers start from an asymmetric position, the density imbalance builds up more rapidly in the shorter arm than in the longer arm. So the dc-to-ac transition occurs earlier when barriers are moved towards the short arm (negative biasing) than towards the long arm (positive biasing), yielding unequal critical currents $|I_\mathrm{c+}| \neq |I_\mathrm{c-}|$. Therefore, by increasing the initial asymmetry, we achieve higher diode efficiency (see Fig.~2 (g)); hence, a position-tunable diode is realized. We further investigate the tunability of diode efficiency by rotating the junctions at a scaled frequency $\Omega'$. Due to the interference of clockwise and counterclockwise currents, critical currents are minimum at half-integer flux~\cite{ryu2020quantum, pradhan2025fractional}, making the diode efficiency minimum as shown in Fig.\ref{fig:fig2} (h).

\textit{Applications:} First, we demonstrate the emergence of asymmetric Shapiro steps in the current-chemical potential difference curve ($f_{\mathrm{bar}}$-$\Delta \omega$ curve) by applying an ac current in the form of periodically driven barriers. In simulation, this motion is modeled as $\phi_{b_{1,2}} = \pm (2 \pi f_{\mathrm{bar}} ) t + \phi_\mathrm{m} \, \sin(2 \pi f_{\mathrm{m}} t)$. Fig.~\ref{fig:fig3} (a) shows the plateau in the chemical potential difference, which is quantized by the external barrier modulation frequency $f_{\mathrm{m}}$, is symmetric under positive and negative barrier velocities for a symmetric configuration of the barriers. In the other case, when the barriers are placed asymmetrically, the Shapiro step heights are quantized, but the location and width of the plateau are different for positive and negative biasing, as seen in Fig.~\ref{fig:fig3} (b). Therefore, for a sufficiently large asymmetry, the Shapiro step can vanish for a particular biasing direction, leading to a nonreciprocal barrier modulation detector. 


We demonstrate the use of the atomtronic Josephson diode as a half-wave rectifier in Fig.~\ref{fig:fig3} (c) and (d). We achieve this functionality by applying a square wave bias current in the form of barrier motion with alternating direction up to $3$ cycles. First, we move the barriers towards the longer arm of the ring with velocity $\pm f_{\mathrm{bar}}$ to induce a positive current, and then reverse the barrier motion with the same velocities to induce the negative bias current. The corresponding current profile is shown in Fig.~\ref{fig:fig3} (c). During the barrier motion period, we keep the cavity field off and turn it on once the barrier motion stops to measure the chemical potential difference. Between successive intervals of turning off and turning on the cavity field, the barriers move for $t_{\mathrm{bar}} = 0.1$ ms, which is long enough for the cavity field to decay (within $\sim \mu$s). Hence, each measurement is independent, not carrying information from the previous measurement. To obtain the rectification, we have used $|f_{\mathrm{c}-}| < |f_{\mathrm{bar}}| < |f_{\mathrm{c}+}|$ and the corresponding chemical potential difference ($\Delta \omega$) for each measurement cycle is shown in Fig.~\ref{fig:fig3} (d). This yields a zero chemical potential difference (`off' state) for the positive current but a non-zero value (`on' state) for the negative current. Hence, the bias current in the form of a square wave is rectified. However, the cycle count is limited by the lifetime of the condensate and the duration of the barrier motion. Additionally, a careful selection of barrier velocity that lies between the critical values is necessary to achieve rectification.


\begin{figure}[t]
\begin{center}
\includegraphics[width= 1\linewidth]{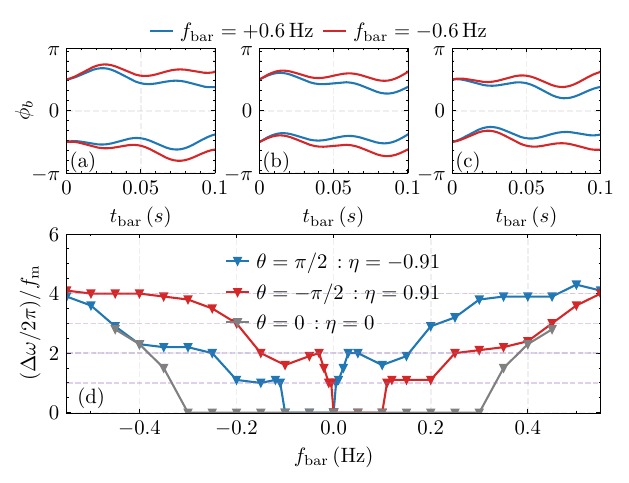} \\

\end{center}
\vspace{-12pt}
\caption{Drive tunable diode: temporal evolution of barrier positions for $f_{\mathrm{m}} =10\,\mathrm{Hz},\,  \phi_\mathrm{m} = 0.22$, (a) $\theta = \pi/2$, (b) $\theta = 0$, (c) $\theta = -\pi/2$. (d) Variation of the magnitude of peak splitting ($\Delta \omega$) versus barrier velocity ($f_{\mathrm{bar}}$). The other set of parameters used here is the same as used in Fig.~\ref{fig:fig2} (a).}
\label{fig:fig4}
\end{figure}

\textit{Drive tunable diode:} We discuss the realization of a Josephson diode by using a biharmonic ac drive. This breaks the spatio-temporal symmetry of the system when the phase difference between the two harmonics $\theta \neq n \pi$, where $n$ is an integer. This phase difference makes the ac drive asymmetric (unequal maximum and minimum), which leads to different barrier motions in forward and backward directions (see Figs.~\ref{fig:fig4} (a) and (c)), resulting in unequal critical currents in the two directions. In simulation, the asymmetric biharmonic ac drive is achieved via modeling the barrier motion as~\cite{tsarev2025fractionalshapirostepsrsj, Borgongino2025biharmonic} 
\begin{equation}
    \phi_{b_{1,2}} = \pm (2 \pi f_{\mathrm{bar}} ) t + \phi_\mathrm{m} \, [\sin(2 \pi f_{\mathrm{m}} t) + \sin(4 \pi f_{\mathrm{m}} t + \theta)].
    \label{eq:acdrive_asym}
\end{equation}

The resulting barrier velocity versus chemical potential difference curve (I-V curve) is shown in Fig.~\ref{fig:fig4} (d), which clearly reveals a difference in the crtical velocity magnitude for the forward $+f_{\mathrm{bar}}$ and backward $-f_{\mathrm{bar}}$ directions, leading to a non-zero diode efficiency of $91 \%$. Additionally, the I-V curve exhibits asymmetric Shapiro steps, a characteristic feature of the JDE~\cite{Borgongino2025biharmonic,Fominov2022AsymSQUID, Souto2022PRL}. As predicted by Eq.~(\ref{eq:acdrive_asym}), reversing the sign of both $f_{\mathrm{bar}}$ and $\theta$ results in identical barrier motion, and cosequently, the same peak splitting magnitude  $\Delta \omega$, in the cavity output spectrum.  This symmetry allows for easy flipping of the diode efficiency's sign by simply changing the sign of $\theta$, while maintaining the same efficiency magnitude. Such a feature enables real-time switching and current rectification with minimal operational requirement.

\textit{Conclusion and future scope:} We have proposed and simulated the realization of the Josephson diode effect in an atomtronic circuit by using asymmetrically placed optical barriers or an asymmetric ac drive. They break the spatio-temporal symmetry of the system, giving rise to unequal critical currents in the forward and backward directions. The diode efficiency is quantified by measuring the difference between the magnitudes of critical currents. We have demonstrated a maximum diode efficiency of $91 \%$, which can be further tuned by the barrier and drive parameters. The capability of this atomtronic Josephson diode is demonstrated through the rectification of a square wave signal, whose rectification cycle can be further increased by reducing the barrier movement and measurement time. Future work will explore the Josephson diode effect in other circuit geometries, including non-Hermitian junctions and with an efficiency of $100\%$. Additionally, the cavity-based measurement technique demonstrates the potential of cavity optomechanics as a single-shot, highly sensitive probe for studying condensate dynamics in atomtronic Josephson circuits.

\textit{Acknowledgements:} We gratefully acknowledge the supercomputing facilities Param Ishan and Param Kamrupa, where all the simulation runs were performed. M.B. thanks the Air Force Office of Scientific Research (FA9550-23-1-0259) for support. R.K. was supported by JSPS KAKENHI (Grant No.25K07190), and JST ERATO (Grant No. JPMJER2302).

\bibliography{citation.bib} 



\clearpage

\widetext

\begin{center}
\textbf{\large Supplementary material: Proposals for realizing a Josephson diode in Atomtronic circuits}
\end{center}

\setcounter{equation}{0} \setcounter{figure}{0} \setcounter{table}{0} %
\setcounter{page}{1} \setcounter{section}{0} \makeatletter
\renewcommand{\theequation}{S\arabic{equation}} \renewcommand{\thefigure}{S%
\arabic{figure}} \renewcommand{\bibnumfmt}[1]{[S#1]} \renewcommand{%
\citenumfont}[1]{S#1} \renewcommand{\thesection}{S\arabic{section}}%
\setcounter{secnumdepth}{3}

\renewcommand{\thefigure}{SM\arabic{figure}} \renewcommand{\thesection}{SM
\arabic{section}} \renewcommand{\theequation}{SM\arabic{equation}}

In this supplementary material, we present a comprehensive set of additional derivations, simulation details, and supporting information to further elucidate the results discussed in the main text. In Section \ref{sec:SMopt}, we provide a detailed derivation of the optical lattice potential that is used within the Gross-Pitaevskii equation to model the system. Section \ref{sec:SMsimdetails} contains an in-depth description of the simulation techniques and parameters employed to generate all the results presented in the main paper. In Section \ref{sec:SMdc_ac}, we demonstrate the  detection of the dc-ac Josephson transition, including the methodology and key observations. Finally, in Section \ref{sec:SMrect}, we present the cavity output spectra that correspond to the rectification of a square wave signal, offering further insight into the rectification process and its impact on the system dynamics.


\section{Optical lattice potential}
\label{sec:SMopt}

The dispersive light-matter interaction, in the limit of large atom-photon detuning, is governed by \cite{pradhan2024cavity, KumarPRL2021}
\begin{align}
    V_{\mathrm{opt}}&\approx\frac{\hbar\mathcal{D}_{eg}^{2}\mathcal{E}_{0}^{2}|u_{+\ell0}(\mathbf{r})+u_{-\ell0}(\mathbf{r})|^{2}}{\Delta_{a}}a^{\dagger}a\;,\label{Coupl1}
\end{align}
where condensate atoms inside the cavity are treated as a two-level system. Here, $\mathcal{D}_{eg}$ is the transition dipole moment operator between the atomic ground and excited states, $a^{\dagger}a$ is the photon number of the cavity mode, $\Delta_{a}$ is the detuning between driving laser frequency and the atomic transition frequency, $\mathcal{E}_{0}$ denotes the single photon electric field amplitude inside the cavity, $u_{\pm \ell0}(\mathbf{r})$ are the mode functions of the Lagurre-Gaussian beams of OAM $\pm \ell\hbar$ and $|u_{+\ell0}(\mathbf{r})+u_{-\ell0}(\mathbf{r})|^{2}$ gives the cavity mode intensity distribution at atomic position $r$. Near $z=0$ plane, the intensity distribution of the superposition is written as $\approx\cos^2(\ell\phi)$. By combining all the constant contributions, we obtain
\begin{align}
 V_{\mathrm{opt}}=\hbar U_{0}\cos^{2}(\ell\phi)a^{\dagger}a\;.\label{Coupl}   
\end{align}

\section{Simulation details}
\label{sec:SMsimdetails}
\begin{figure*}[!htb]
\begin{center}
\includegraphics[width= 1\linewidth]{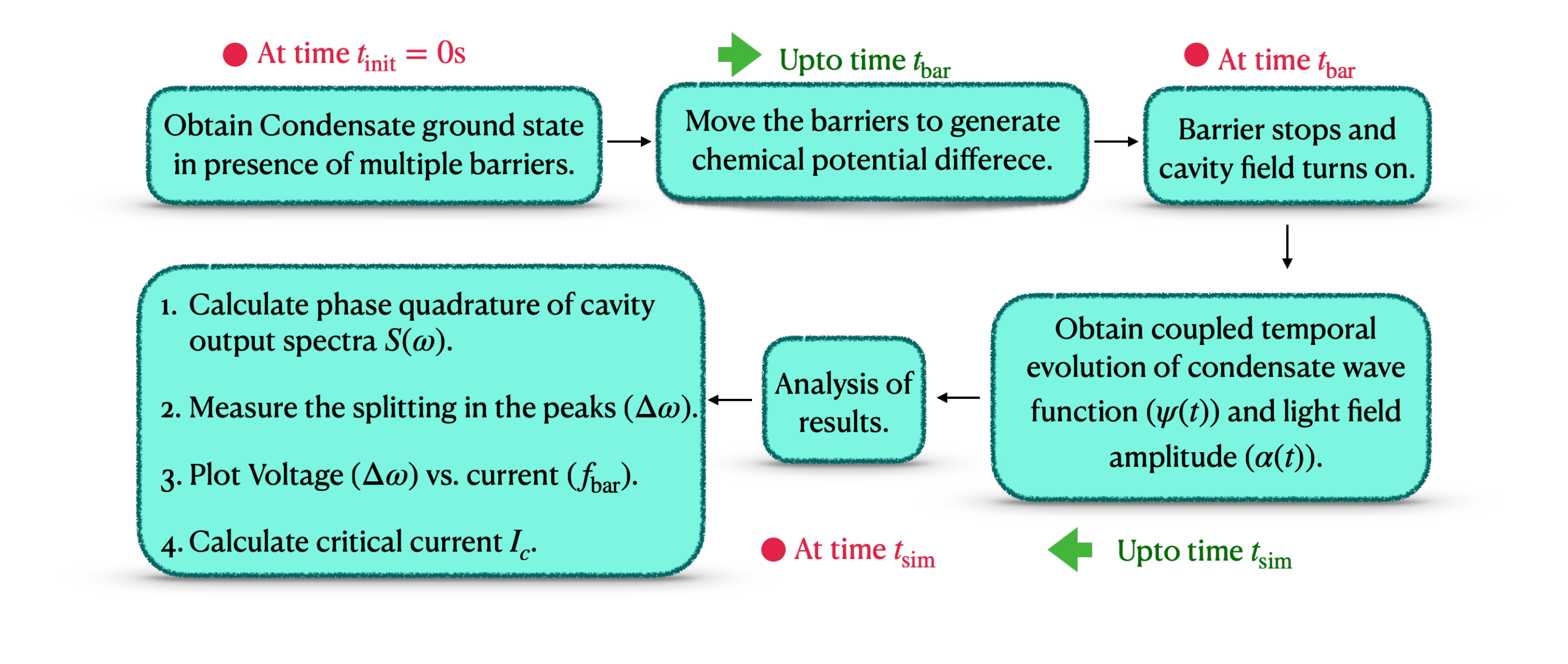} \\

\end{center}
\vspace{-12pt}
\caption{ Flowchart of simulation details.}
\label{fig:figSM1}
\end{figure*}

In this section, we present the details of all the simulations conducted in this work as a flowchart in Fig.~\ref{fig:figSM1}.

\section{Detection of dc-ac transition and Josephson oscillation frequency}
\label{sec:SMdc_ac}

\begin{figure*}[b]
\begin{center}
\includegraphics[width= 1\linewidth]{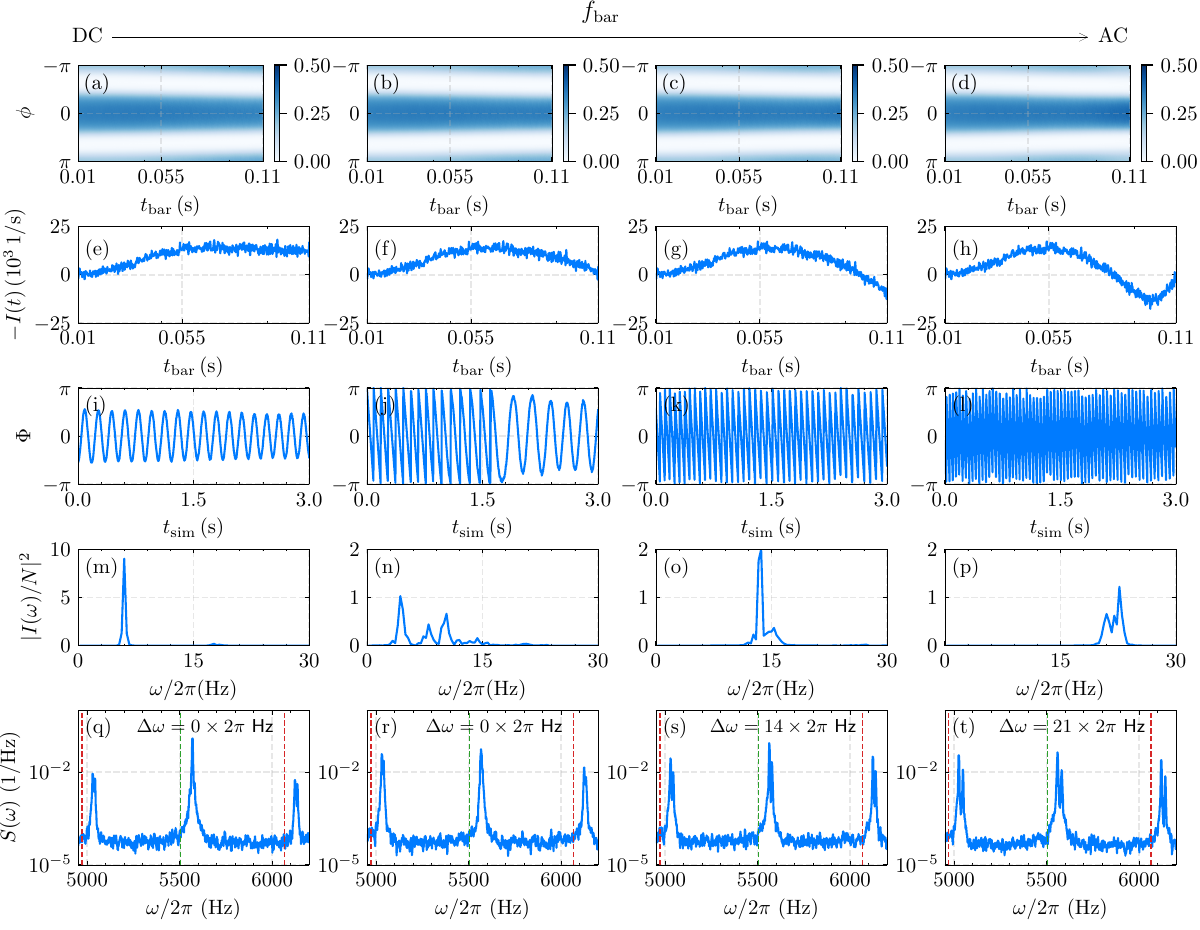} \\

\end{center}
\vspace{-12pt}
\caption{DC-AC Josephson transition [(a)-(d)]. Time evolution of condensate density during the barrier movement. [(e)-(h)] Time evolution of particle current during the junction movement. [(i)-(l)] Evolution of phase difference across the barrier during the measurement period. [(m)-(n)] Power spectra of population imbalance during the measurement period. [(q)-(t)] Power spectra of phase quadrature of the cavity output field vs. the system's response frequency. The green and red vertical dashed lines indicate the analytical prediction of side mode frequencies of a uniform density ring-condensate corresponding to $L_p = 0$ and $L_p = 1$, respectively, obtained from the BdG analysis (Eq.~\ref{eq:OmP}). Here, junction velocities are $f_{\mathrm{bar}} = 0.3 \, \mathrm{Hz},\, 0.33 \, \mathrm{Hz}, \, 0.35 \, \mathrm{Hz}$, and $ 0.4 \, \mathrm{Hz}$ respectively. The other set of parameters is the same as used in Fig.~\ref{fig:fig2} (b).}
\label{fig:figSM2}
\end{figure*}

In order to characterize the dc-ac Josephson regime, we illustrate the simulation results for increasing barrier velocities $f_{\mathrm{bar}}$ in Fig.~\ref{fig:figSM2}, which include the temporal evolution of condensate density (first row), tunnelling current (second row) during the barrier movement period, temporal evolution of phase difference between the two wells (third row), power spectrum of tunneling current (fourth row), and the power spectrum of the phase quadrature of cavity output field during the cavity measurement period, once the barrier motion stops (fifth row). 

For small barrier velocities $f_{\mathrm{bar}}$, barrier motion induces a flow of atomic supercurrent, maintaining the same atomic density between the two regions~\cite{levy2007acdc,RyuPRL2013} (see Figs.~\ref{fig:figSM2} (a)-(b)). But, when $f_{\mathrm{bar}}$ is increased beyond a critical value $f_{\mathrm{c}}$, we see the compression of atoms in the region towards which the barriers move (see Figs.~\ref{fig:figSM2} (c)-(d)). This results in a non-zero chemical potential difference between the two regions $\Delta \mu$, giving rise to the Josephson oscillation of atomic current (see Figs.~\ref{fig:figSM2} (g)-(h)), which is otherwise a constant dc current for lower barrier velocities (see Fig.~\ref{fig:figSM2} (e)). For barrier velocity $f_{\mathrm{bar}} = 0.33$ Hz (see Fig.~\ref{fig:figSM2} (f)), the atomic current just starts to oscillate, marking the onset of transition from the dc to ac regime.

This dc-ac transition point is further confirmed by calculating the temporal evolution of the phase difference $\Phi$ across the two regions, which are shown in Figs.~\ref{fig:figSM2} (i)-(l). For small barrier velocities, the phase difference undergoes bounded, small-amplitude oscillations that are induced by the barrier. As we increase the barrier velocity, the phase difference starts increasing monotonically, entering the running-phase mode. This temporal evolution of the phase difference obeys the relation~\cite{levy2007acdc,experiment1DJJ}
\begin{equation}
    \frac{d\Phi}{dt} = -\frac{\Delta\mu}{\hbar},
\end{equation}
which indicates the system has entered the AC Josephson regime. 

Next, we extract the oscillation frequency from the power spectral density (PSD) of the temporal evolution of the tunnelling current for the four cases and show in Figs.~\ref{fig:figSM2} (m)-(p). At the critical barrier velocity, the detection of oscillation frequency is ambiguous, since the phase slips are irregular (see Figs.~\ref{fig:figSM2} (j) and (n)). But beyond the critical value, when the chemical potential difference $\Delta\mu$ is sufficiently large, PSD yields a clear value of the oscillation frequency ($\omega_\mathrm{J} = \frac{\Delta\mu}{\hbar}$), which further increases with increasing barrier velocity (see Figs.~\ref{fig:figSM2} (o)-(p)). This oscillation frequency $\omega_\mathrm{J}/2 \pi$ is also equal to the number of $2 \pi$ crossings of the phase difference per second, as seen by comparing Figs.~\ref{fig:figSM2} (o)-(p) with Figs.~\ref{fig:figSM2} (k)-(l). 

While the observables mentioned above indicate a clear transition from the DC to AC Josephson regime, in our proposed experimental method, the accessible quantity is the power spectrum of the phase quadrature of the cavity output field, illustrated in Figs.~\ref{fig:figSM2} (q)-(t). For small barrier velocities, we obtain the dominant side mode peak corresponding to winding number states $L_p = 0$ (initial winding number of the condensate before introducing the barriers) around $5600 \times 2 \pi$ Hz and two other modes corresponding to $L_p = \pm 1$, those induced by the barrier motion. These side-mode peaks in the cavity output spectrum can be identified from the Bogoliubov-de Gennes (BdG) analysis for a uniform-density ring condensate as~\cite{KumarPRL2021, pradhan2024ring}
\begin{equation}
\label{eq:OmP}
\omega_{c,d}'=\left[\omega_{c,d}\left(\omega_{c,d}+4\tilde{g}N\right)\right]^{1/2},   
\end{equation}
where $\tilde{g}=g/(4\pi\hbar)$, $g$ is the reduced atomic interaction strength, $N$ is the condensate atom number and
\begin{align}
    \omega_{c,d}&=\omega_{\beta}\left(L_p\pm 2\ell\right)^{2}\;,\label{Eq:Freq}
\end{align}
where $\ell$ is the OAM of the Laguerre-Gaussian beams that probe the condensate. 


\begin{figure*}[b]
\begin{center}
\includegraphics[width= 0.6\linewidth]{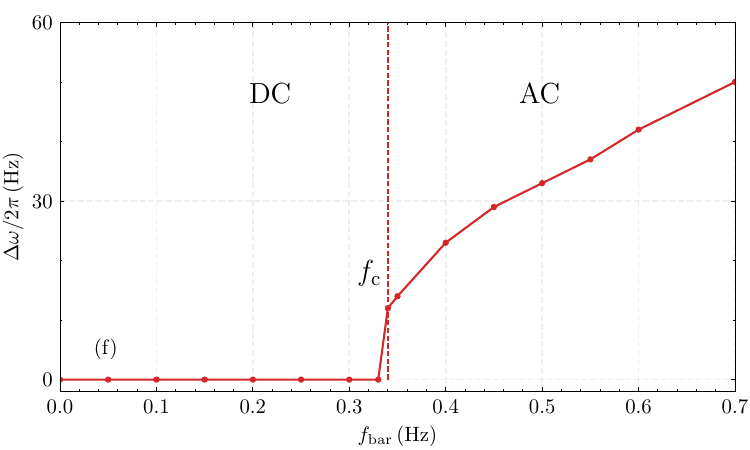} \\

\end{center}
\vspace{-12pt}
\caption{DC-AC Josephson transition: Magnitude of splitting of the side mode peaks in the cavity output spectra $\Delta \omega$ versus barrier velocity $f_{\mathrm{bar}}$. The vertical dashed line indicate the critical barrier velocity beyond which $\Delta \omega \neq 0$.  The other set of parameters is the same as used in Fig.~\ref{fig:fig2} (b).}
\label{fig:figSM3}
\end{figure*}

For barrier velocities greater than the critical value, all the peaks further split by the same amount as a result of the population imbalance oscillation, and the magnitude of the split $\Delta \omega$ is equal to the oscillation frequency, as demonstrated in Figs.~\ref{fig:figSM2} (s) and (t). Hence the Josephson oscillation frequency for each barrier velocity can be calculated by measuring the amount of split of the side-mode peaks in the cavity output spectra. Additionally, the onset of the dc-ac transition can be detected by plotting the magnitude of peak splitting $\Delta \omega$  versus barrier velocity $f_{\mathrm{bar}}$, and the critical barrier velocity is assigned to the value at which the splitting $\Delta \omega$ starts to increase by increasing the barrier velocity $f_{\mathrm{bar}}$, as shown in Fig.~\ref{fig:figSM3}. 


\section{Demonstration of half-wave rectification using position-tunable diode}
\label{sec:SMrect}

\begin{figure*}[!htb]
\begin{center}
\includegraphics[width= 1\linewidth]{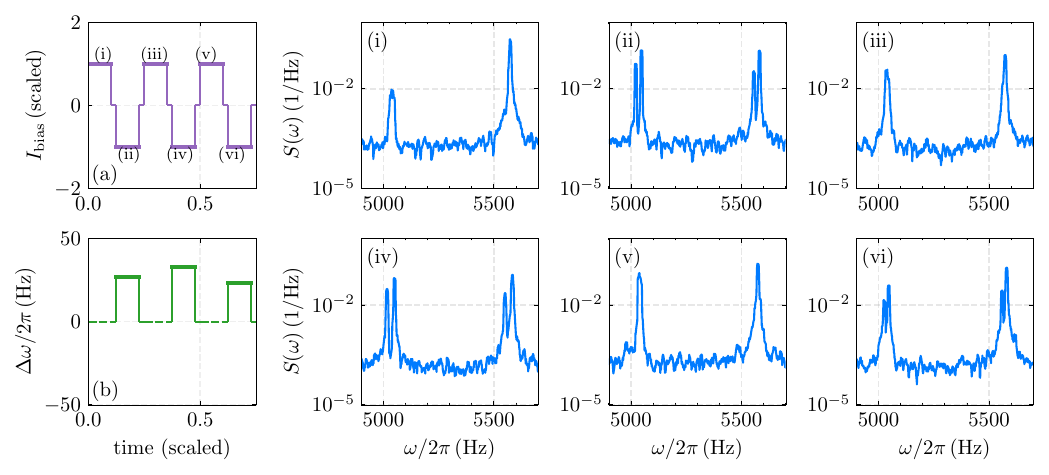} \\

\end{center}
\caption{ Half-wave rectifier. (a) Three cycles of square wave of bias current that corresponds to $f_{\mathrm{bar}} = \pm 0.26$ Hz (lies in between $f_{\mathrm{c}+}$ and $f_{\mathrm{c}-}$ ). (b) Measured junction voltage ($\Delta \omega$). (c) Cavity output spectra of corresponding $\pm f_{\mathrm{bar}}$. The set of parameters used is the same as used in Fig.~\ref{fig:fig2} (b).}
\label{fig:figSM4}
\end{figure*}

In this section, we present the cavity output spectra for the case of half-wave rectification using a position-tunable diode.  A square wave bias current signal is generated by moving the barriers towards each other up to time $t_{\mathrm{bar}}$ with velocity $f_{\mathrm{bar}}$ , keeping them static during the measurement period, and then moving them towards each other in the opposite direction with velocity $-f_{\mathrm{bar}}$. The barrier motion in positive and negative barriers generates the bias currents of opposite polarity, as shown in Fig.~\ref{fig:figSM4} (b). The cavity field is turned on after each barrier movement period (once the barrier motion stops) to measure the chemical potential difference or the Josephson oscillation frequency. The corresponding cavity output spectra for three current cycles are shown in Figs.~\ref{fig:figSM4} (i-vi). We have extracted the amount of split from each measurement interval and plotted it versus time in Fig.~\ref{fig:figSM4} (b).  The alternating positive and negative bias current yields zero and a non-zero amount of splitting of the side mode peaks, respectively, thereby realizing a half-wave rectifier. Ideally, the magnitude of the non-zero split should remain the same for equal barrier velocities. However, in our simulation, they differ slightly. This might arise due to the hysteresis in the condensate response or residual excitation from the previous cycle.




\end{document}